\documentclass[a4paper,11pt]{article}
\usepackage{graphicx,rotating,wrapfig,float}
\usepackage{units}
\usepackage[pdftex]{hyperref}
\hypersetup{colorlinks,linkcolor=blue}
\usepackage[margin=10pt,font=small,format=plain,labelfont=bf]{caption}
\usepackage{enumerate}
\usepackage{amsmath}
\usepackage{authblk}
\usepackage{subfigure}
\usepackage{tabularx}

\usepackage{epstopdf}
\newcommand{\ardm}{ArDM}

\title{\bf Status of \ardm-1t: First observations from operation with a full ton-scale liquid argon target\\}
\author[1]{J.~Calvo}
\author[1]{C.~Cantini}
\author[2]{M.~Daniel}
\author[1]{U.~Degunda}
\author[1]{S.~Di~Luise}
\author[1]{L.~Epprecht}
\author[1]{A.~Gendotti}
\author[1]{S.~Horikawa}
\author[1]{L.~Knecht}
\author[2]{B.~Montes}
\author[1]{W.~Mu}
\author[1]{M.~Munoz}
\author[1]{S.~Murphy}
\author[1]{G.~Natterer}
\author[1]{K.~Nguyen}
\author[1]{K.~Nikolics}
\author[1]{L.~Periale}
\author[1]{C.~Regenfus}
\author[2]{L.~Romero}
\author[1]{A.~Rubbia\thanks{andre.rubbia@cern.ch}}
\author[2]{R.~Santorelli}
\author[1]{F.~Sergiampietri}
\author[1]{D.~Sgalaberna}
\author[1]{T.~Viant}
\author[1]{S.~Wu}

\affil[1]{ETH Zurich, Institute for Particle Physics, CH-8093 Z\"{u}rich, Switzerland}
\affil[2]{CIEMAT, Div. de F{\'\i}sica de Particulas, Avda. Complutense, 22, E-28040, Madrid, Spain}

\date{{\bf\small The \ardm\ Collaboration\\}
\vspace{3mm}
\small{May 10, 2015\vspace{-1mm}}}

\begin{document}
\maketitle

%\vspace{-2mm}
\begin{abstract}
{\ardm\ is the first operating ton-scale liquid argon detector for direct search of Dark Matter particles. Developed at CERN as Recognized Experiment RE\,18, the experiment has been approved in 2010 to be installed in the Spanish underground site LSC (Laboratorio Subterraneo de Canfranc). Under the label of LSC EXP-08-2010 the \ardm\ detector underwent an intensive period of technical completion and safety approval until the recent filling of the target vessel with almost 2\,t of liquid argon. This report describes the experimental achievements during commissioning of \ardm\ and the transition into a stage of first physics data taking in single phase operational mode (\ardm\,Run\,I).  We present  preliminary observations from this run. A first indication for the background discrimination power of LAr detectors at the ton-scale is shown. 
We present an outlook for completing the detector with the electric drift field and upgrade of the scintillation light readout system with novel detector modules based on SiPMs in order to improve the light yield.}
\end{abstract}

\vspace{-2mm}

%\newpage
%\tableofcontents

\textwidth=16truecm    

%%%%%%%%%%%%%%%%%%%%%%%%%%%%%%%%%%%%%%%%%%%%%%%%%%
\section{Introduction}
\label{sec:intro}
%%%%%%%%%%%%%%%%%%%%%%%%%%%%%%%%%%%%%%%%%%%%%%%%%%

In February 2015 the \ardm\ experiment~\cite{Rubbia:2005ge, Mar:2011, Bad:2013} achieved a major milestone by completing the filling of the detector vessel with a total of nearly 2\,t of liquid argon (LAr). Now the project has entered the first period of physics data taking in the single-phase operation mode (\ardm\,Run\,I). This paper is based on a report submitted to the scientific committee of LSC in April 2015 and presents recent experimental accomplishments of \ardm , including the status of the underground operation at LSC, the progress in data taking and analysis, as well as the in-situ measurement of the environmental neutron flux in Hall A. Emphasis is also put on the mid-term future describing for the first time potential upgrades of ArDM. 

The commissioning of the detector went smoothly and is described in detail in chapter\,\ref{sec:filling}. The next step in the experimental program is to complete the external neutron shield with the missing tiles from the top cover which were not yet installed for better access to the main vessel during the filling period. After we plan to continue mass data taking in the single-phase operation mode through June 2015. This will provide a precise assessment of the long-term stability of the system by regular calibration runs with radioactive test sources. Having achieved these major milestones, \ardm\ proposes to evolve in the near future with detector upgrades and a second period of physics data taking in the double-phase TPC operation mode. 

%\ardm\ was recently reconfirmed as a CERN Recognised Experiment (RE18) until 2018, with possible third renewal. This allows for the continued access to CERN infrastructure and technical staff, which is essential for the remote operation of \ardm\ from our \ardm\ Control Centre at CERN, and the use of CERN facilities (like CASTOR) for the storage of the data. 

%Since December 2014 we are also in close contact with the DarkSide group, which on their part operate a 50\,kg LAr double phase detector at LNGS\,\cite{ds50}. This collaboration is underpinned by regular meetings to coordinate mutual support for upgrades of both, the \ardm\ and Darkside experiments, as well as the R\&D efforts towards a large LAr based common future project. 

%%%%%%%%%%%%%%%%%%%%%%%%%%%%%%%%%%%%%%%%%%%%%%%%%%
\section{Status of the underground operation at LSC}
\label{sec:status-exp}
%%%%%%%%%%%%%%%%%%%%%%%%%%%%%%%%%%%%%%%%%%%%%%%%%%
\subsection{Filling of the detector vessel with 2\,t of liquid argon}
\label{sec:filling}
%%%%%%%%%%%%%%%%%%%%%%%%%%%%%%%%%%%%%%%%%%%%%%%%%%

\begin{figure}[H]
\begin{center}
\includegraphics[width=0.91\columnwidth]{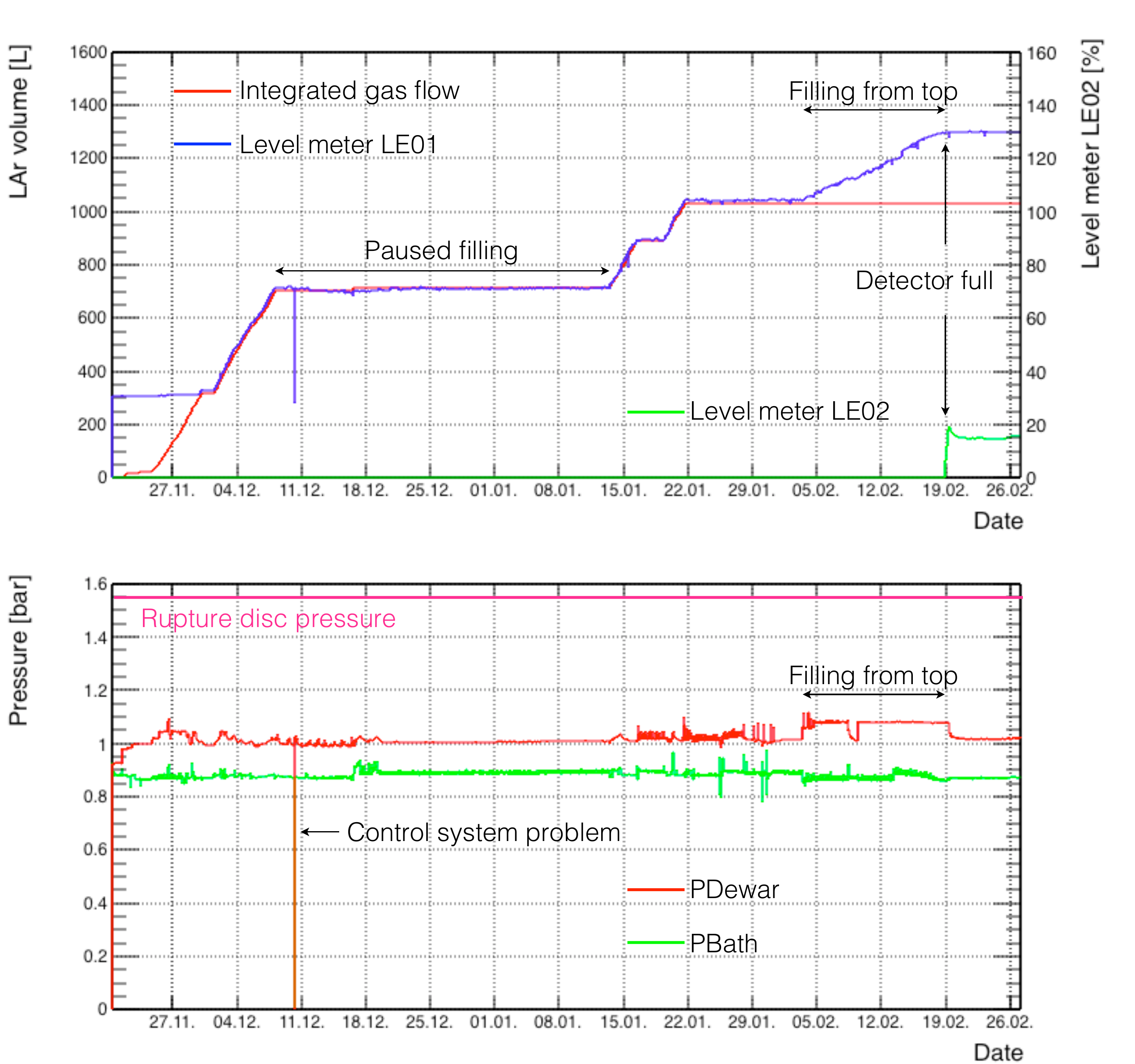}
\caption{Entire history of the filling of the \ardm\ detector vessel. Top: the amount of LAr condensed into the system, evaluated in two different ways. The red curve shows the calculated amount of LAr from the integrated gas flow recorded using a flow meter. The blue curve shows the liquid level inside the detector vessel measured with a $\sim$\,1.5\,m long cylindrical level meter (LE01). The liquid level was calibrated to the LAr volume to match the red curve. LE01 started to show a reading after $\sim$\,300\,$\ell$ had been condensed in. Above $\sim$\,1050\,$\ell$ the flow meter could not be used for monitoring since the filling path was changed (see text). The green curve shows the values of one of the three small level meters (100\% for 20\,mm in height) mounted around the nominal LAr level indicating the completion of the filling. Bottom: Pressures in the detector vessel (PDewar, red) and in the cooling argon bath volume (PBath, green) for the same period as in the top plot. The large distance to the safety value demonstrates the large safety margin of the cryogenic system.}
\label{fig:filling}
\end{center}
\end{figure}

After an extensive commissioning period of the experimental installation by means of a cold gaseous argon target, the \ardm\ detector was embedded in the safety infrastructure of LSC. The final permission to fill the detector vessel with liquid argon was granted in November 2014 and filling of the main detector vessel was started shortly after. Figure\,\ref{fig:filling} displays the entire history of the filling period over a time interval of roughly three months with interruptions due to technical improvements or public holidays. The filling was achieved by condensing gaseous argon (purity 99.9999\%\footnote{ALPHAGAZ 2 grade argon from Air Liquide - Spain.}) from banks of 200\,bar gas bottles into the system using the three installed 300\,W Gifford-McMahon cryocoolers. A total of $\sim$\,6.5 banks of 16$\times$50\,$\ell$ gas bottles containing 1040 standard m$^3$ of argon gas, were consumed to fill the vessel with $\sim$\,1300\,$\ell$  of LAr. 

The top plot in Fig.\,\ref{fig:filling} shows the time evolution of the accumulated LAr volume in litres sent into the system. The condensed amount of LAr was controlled continuously in three ways: (1) gas-bottle pressure, (2) integrated gas flow record by a flow meter (red curve) and (3) a $\sim$\,1.5\,m long cylindrical capacitive level meter (LE01, blue curve) installed inside the detector vessel. 
The bottom of LE01 is located approximately at the top surface of the bottom PMTs, so a first reading could be recorded only when the system already contained $\sim$\,300\,$\ell$ of liquid. The flow meter could not be used during the last part of the filling when the vessel had to be supplied through the top flange. The blue curve (LE01) in Fig.\,\ref{fig:filling} was calibrated to match the red curve up to the level of $\sim$\,1050\,$\ell$. 

The \ardm\ cryocoolers (Cryomech AL300) provide a nominal cooling power of 266 W at 80~K with 380\,V 50\,Hz AC power. With a measured total heat load of $\sim$\,400\,W into the \ardm\ system, approximately 400\,W of cooling power is available for argon liquefaction when all three cryocoolers operate in parallel. Taking into account the heat of cooling argon gas from room temperature to LAr temperature (87\,K) and the latent heat of liquefaction, a maximum filling rate was estimated to 90\,$\ell$ of LAr per day, corresponding to 14.4\,d filling time for the full 1300\,$\ell$. 
A constant filling rate of 50--70\,$\ell$/d was achieved consistently resulting in approximately five weeks of actual filling. Several short intervals for exchanging the gas-bottle banks and two long pauses lead to a three months real filling time.
For most of the time argon was condensed in a cooling coil connected to the bottom of the vessel. The cooling coil is completely immersed  in the cooling argon bath volume and has the most efficient heat exchange with the cryocoolers. 
For the last 20\% of volume we opted for filling from the top flange of the vessel, for a safer filling to cope with the hydrostatic pressure present at the bottom of the vessel. 

The completion of the filling was controlled by three small capacitive level meters.  They are mounted around the position of  the nominal liquid level that is $\sim$\,50\,mm below the bottom end of the top PMTs, and are capable of measuring the liquid level in sub-millimetre precision over their entire height of 20 mm. The values from one of the three (LE02, green curve) are shown in the upper part of Fig.\,\ref{fig:filling}. 
The filling was completed on 18 February 2015.
The pressures inside the detector vessel (PDewar, red curve) and the cooling bath (PBath, green curve) are shown in the bottom plot of Fig.\,\ref{fig:filling}. During filling the vessel pressure was limited to below $\sim$\,1.1\,bar (absolute), which is still far below the rated pressure of the vessel and the rupture pressure of the safety discs. For most of the time it was kept below 1050\,mbar, while during the last part of the filling the pressure was raised to 1.1\,bar to improve the condensing efficiency. The bath pressure was kept 0.1--0.2\,bar below the vessel pressure, to keep the LAr temperature in the cooling bath lower by 1--2\,K with respect to the inside of the vessel for efficient heat exchange. Once the detector vessel was fully filled, the total heat input increased from 400\,W to approximately 470\,W. 

Since the beginning of February 2015, we have been operating the LAr recirculation pump continuously. The decay time of the slow scintillation component is used as measure for our liquid purity. This method not sensitive to oxygen-equivalent impurities below ppm levels, so the effect of the purification could only be observed on a small range. The slow component decay time has been maintained very constant for a long period of the liquid operation. It has also been observed that the stability of the thermodynamic conditions in the \ardm\ system improves with the LAr recirculation.

%%%%%%%%%%%%%%%%%%%%%%%%%%%%%%%%%%%%%%%%%%%%%%%%%%
\subsection{The \ardm\ Control Centre at CERN}
\label{sec:ardmccc}
%%%%%%%%%%%%%%%%%%%%%%%%%%%%%%%%%%%%%%%%%%%%%%%%%%

\begin{figure}[hbtp]
\begin{center}
\includegraphics[width=0.66\columnwidth]{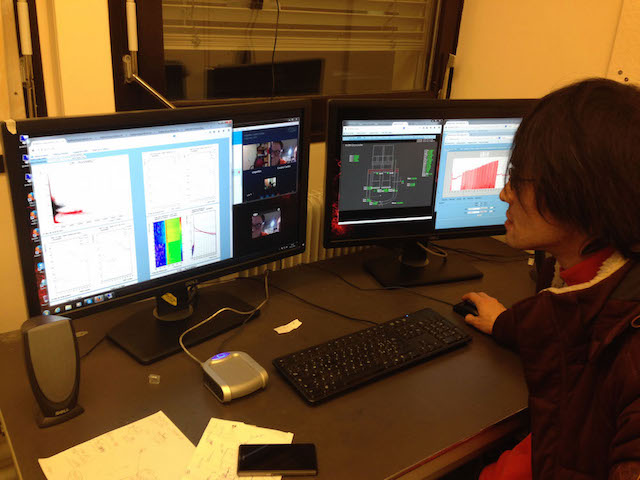}
\caption{The \ardm\ Control Centre at CERN.}
\label{fig:ArDMCCC}
\end{center}
\end{figure}

Since beginning of December 2014 we have set up the main \ardm\ Control Centre at CERN shown by the picture of Fig.\,\ref{fig:ArDMCCC}. Thanks to the full detector control system based on PLC (Programmable Logic Controller), full remote control is possible over the internet. The data acquisition, as well as the online data processing system running locally at LSC,  can also be controlled remotely.
When necessary, interventions at the experimental site are seconded by shift personnel present in the Control Centre via a video conferencing link.

%%%%%%%%%%%%%%%%%%%%%%%%%%%%%%%%%%%%%%%%%%%%%%%%%%
\section{Status of data taking in liquid argon and analysis}
\label{sec:status-daq}
%%%%%%%%%%%%%%%%%%%%%%%%%%%%%%%%%%%%%%%%%%%%%%%%%%

\ardm\ has been running in data acquisition (DAQ) mode continuously and data are recorded  to our 192-TB RAID data storage system installed at LSC. Taking data was also during the  filling period. 
This allowed the monitoring of the system's stability like e.g.~liquid purity, and recording of the evolution of the trigger rate which was increasing with the volume of the LAr target, as expected form the intrinsic activity of atmospheric argon. Several calibration measurements were taken with the approximately half-filled detector, using an external $^{57}$Co source and injecting internally the $^{83{\rm m}}$Kr source. We are now in  physics data taking mode in the single-phase operation. 

%%%%%%%%%%%%%%%%%%%%%%%%%%%%%%%%%%%%%%%%%%%%%%%%%%
\subsection{Data taking and data transfer status}
\label{sec:daq}
%%%%%%%%%%%%%%%%%%%%%%%%%%%%%%%%%%%%%%%%%%%%%%%%%%

\begin{figure}[hbtp]
\begin{center}
\includegraphics[width=0.85\columnwidth]{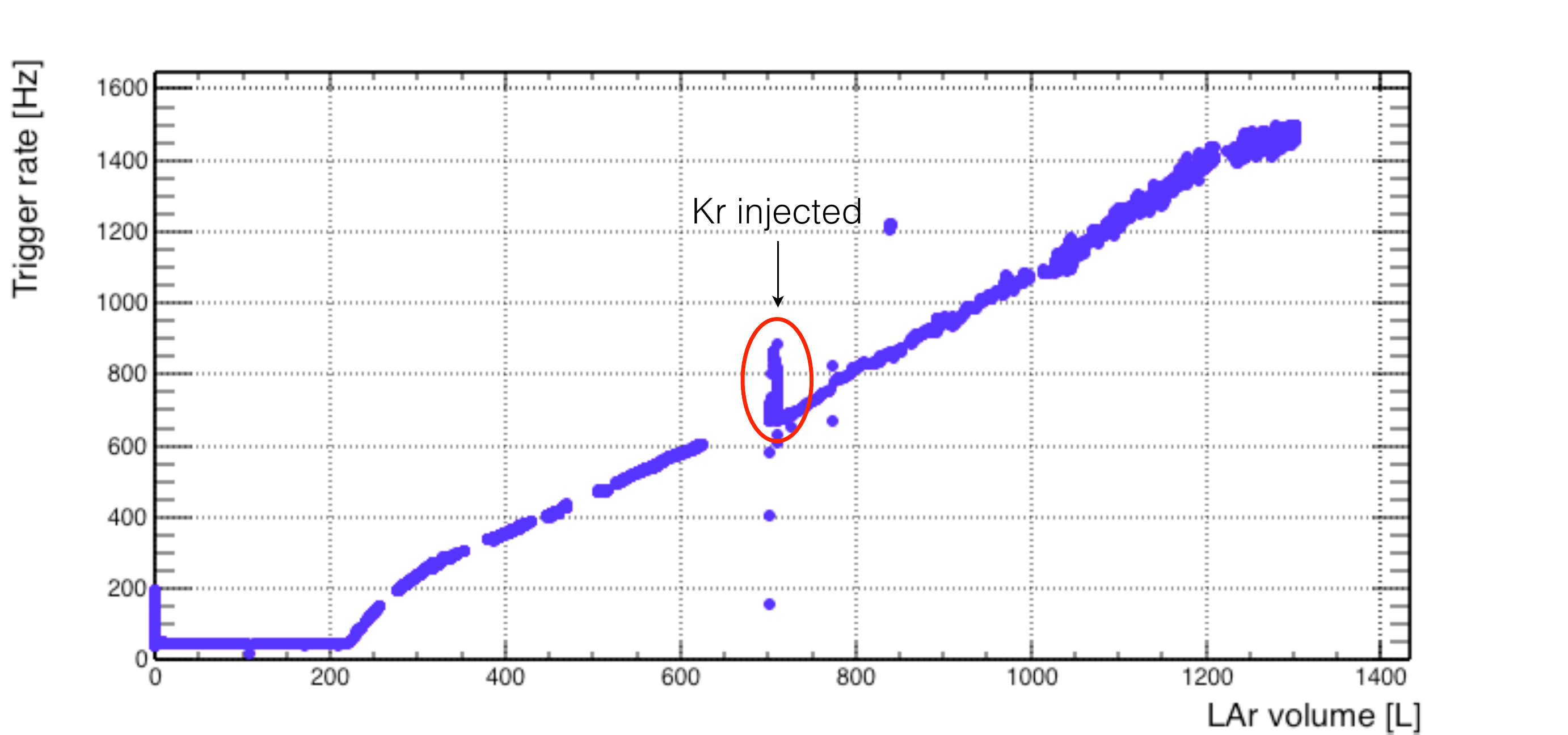}
\caption{Trigger rate as function of the LAr volume dominated by decays
of  $^{39}$Ar isotopes. The trigger signal was generated by logic AND of signals ($\ge$2\,pe each) from the two PMT arrays. The Increase of the trigger rate by the injection of the $^{83{\rm m}}$Kr source can be also be observed (red ellipse).} % Threshold was set to $\sim$\,2 p.e. on each array.}
\label{fig:trgcorr}
\end{center}
\end{figure}

The trigger rate with the detector full of atmospheric argon is at the kHz scale, dominated by decays
of  $^{39}$Ar isotopes. During the filling of the detector with LAr, an approximately linear increase of the trigger rate with the condensed amount of LAr inside the system was observed as shown in Fig.\,\ref{fig:trgcorr}. 

The \ardm\ data acquisition (DAQ) system is designed to comply with this trigger rate, which leads to a data rate of $\sim$\,100 MB/s. The expected large amount of data allows for a very high statistic study of all relevant parameters, with of the order of 10$^8$ events recorded per day. 

\begin{figure}[hbtp]
\begin{center}
\includegraphics[width=0.8\columnwidth]{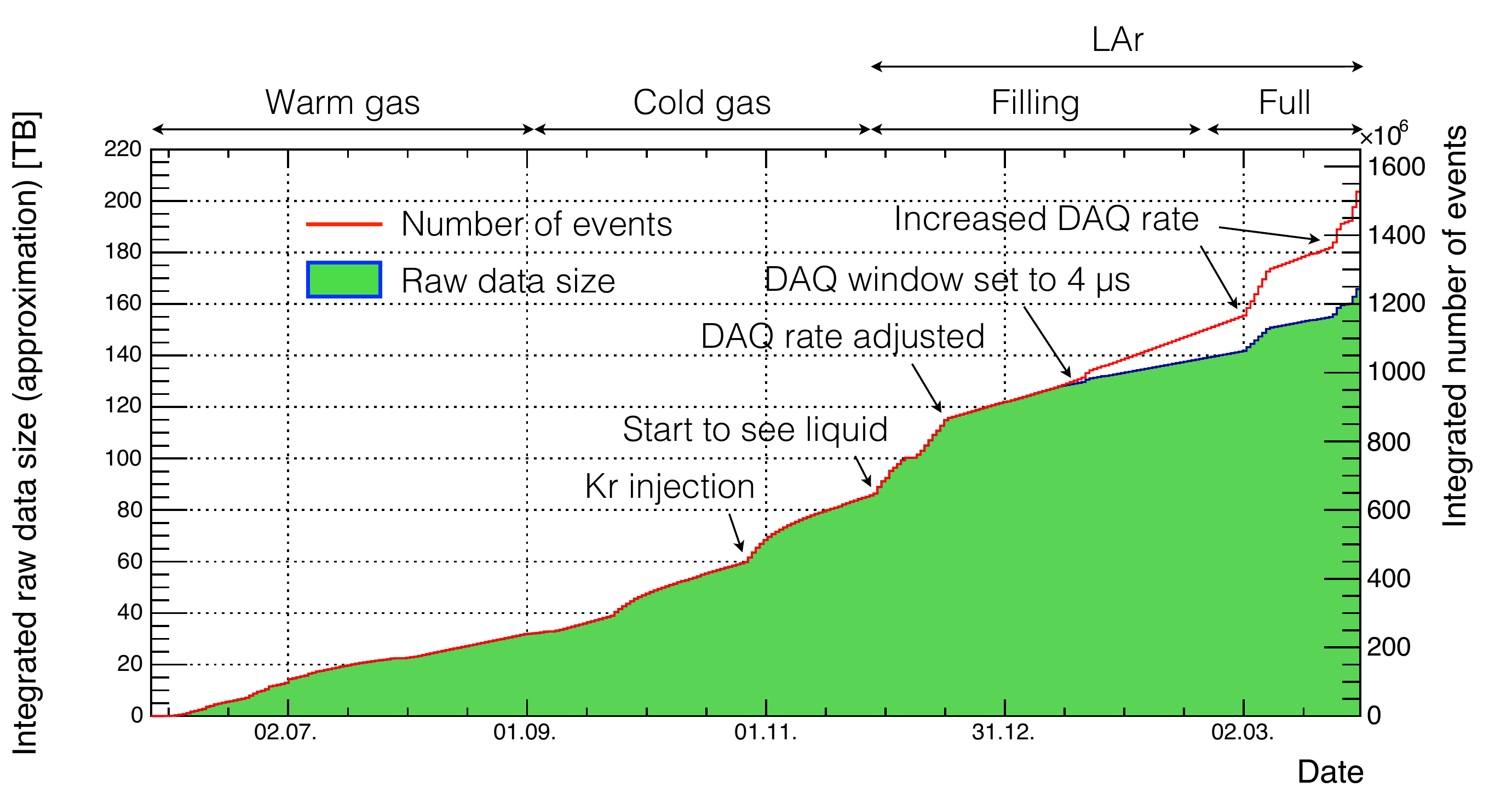}
\caption{Time evolution of integrated size of the raw data that has been acquired (green area) and of the number of recorded events (red curve). Since the middle of the filling period, DAQ rate has been adjusted by pausing the DAQ after each run (90k triggers) so that the data transfer to CERN can keep up with the data taking rate. Also the acquisition window has been reduced to 4 $\mu$s (from 8 $\mu$s) for better matching with the decay time of slow scintillation component in LAr.}
\label{fig:datasize}
\end{center}
\end{figure}

Figure\,\ref{fig:datasize} shows the amount of raw data and the accumulated number of recorded events. The data size at the moment of writing is 160\,TB, resulting in 40\,TB after lossless compression. 

%%%%%%%%%%%%%%%%%%%%%%%%%%%%%%%%%%%%%%%%%%%%%%%%%%%
%%%%%%%%%%%%%%%%%%%%%%%%%%%%%%%%%%%%%%%%%%%%%%%%%%%
%\input{SC_Report}
%\section{Analysis of Liquid Argon Data}
\subsection{Calibration with radioactive sources}
\begin{figure}[!htp]
\centering
\includegraphics[width=0.485\textwidth]{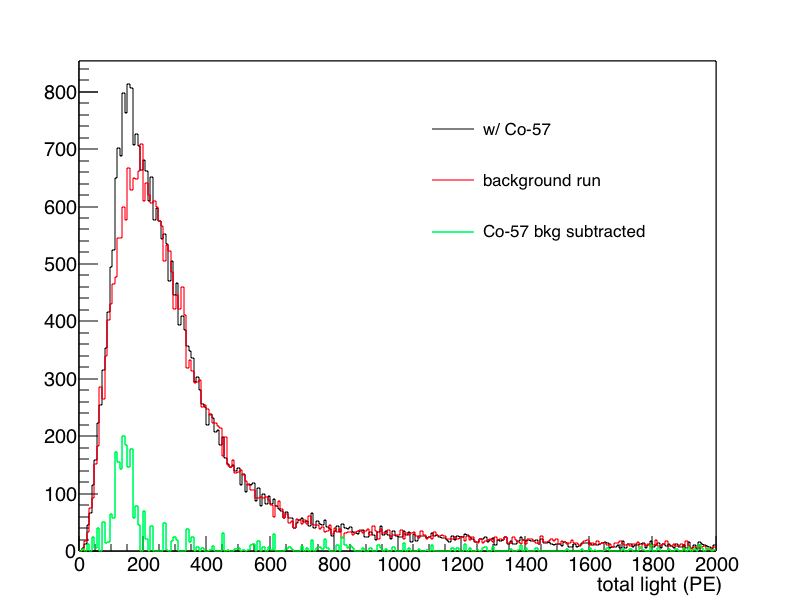}
\includegraphics[width=0.485\textwidth]{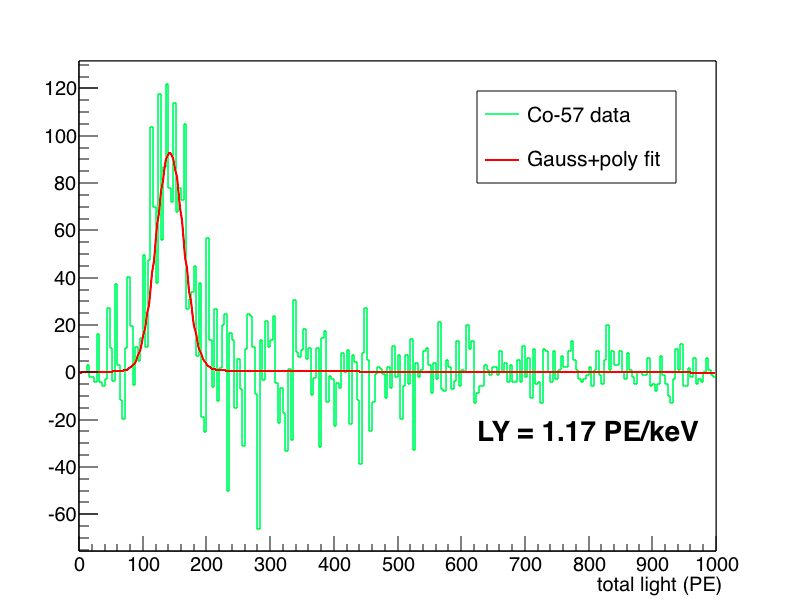}
\caption{Calibration run with the $^{57}$Co source with an energy of 122 keV taken in LAr. (Left) The spectrum of the $^{57}$Co source (black) compared to %a 
that of a background run (red) %in 
is shown. (Right) The Gaussian fit to the calibration peak yields a value for the LY of 1.17 PE/keV$_{\rm ee}$.}
\label{fig:Co57LAr}
\end{figure}

\begin{figure}[!htp]
\centering
\includegraphics[width=0.485\textwidth]{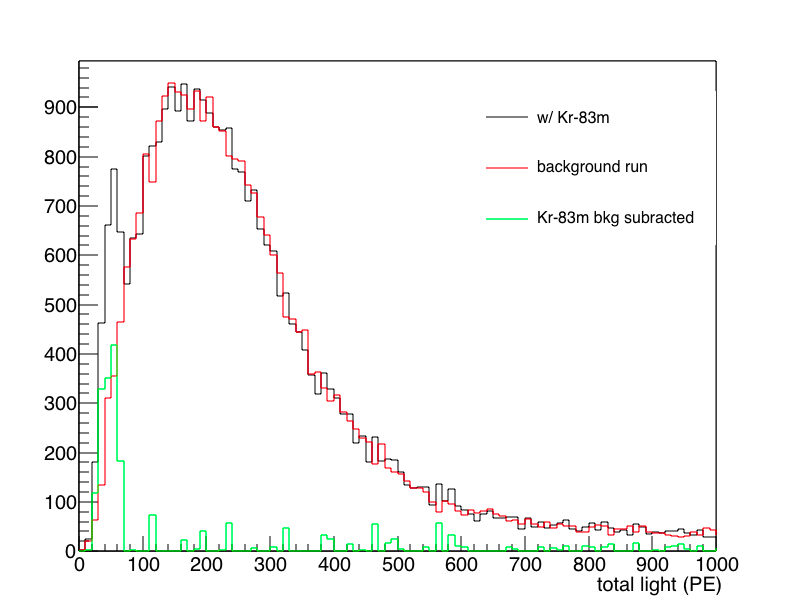}
\includegraphics[width=0.485\textwidth]{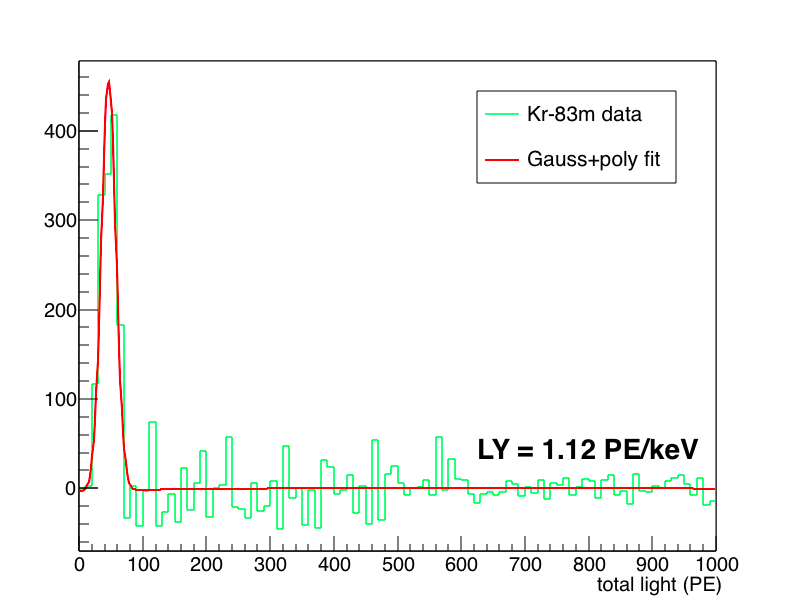}
\caption{Calibration run with the $^{\rm 83m}$Kr source with an energy of 41.5 keV in LAr. (Left) The spectrum of the source (black) compared to a background run (red) clearly shows the position of the calibration peak. (Right) The calibration peak %if 
is fitted with a Gaussian. From the fit we obtain a LY value of 1.12 PE/keV$_{\rm ee}$.}
\label{fig:Kr83mLAr}
\end{figure}

Several calibration runs with the $^{57}$Co (122-keV line) and the $^{83{\rm m}}$Kr (41.5-keV line) sources were acquired to evaluate the light yield (LY) of the detector filled with LAr approximately to the half of the full volume. Figure\,\ref{fig:Co57LAr} shows the spectrum and the fit to the peak from $^{57}$Co, giving a value for LY of 1.17 photoelectrons (PE)/keV$_{\rm ee}$. Figure\,\ref{fig:Kr83mLAr} shows the same for $^{83{\rm m}}$Kr, the LY of 1.12 PE/keV$_{\rm ee}$ from the fit being in good agreement with the value from $^{57}$Co. 

%\clearpage 

%%%%%%%%%%%%%%%%%%%%%%%%%%%%%%%%%%%%%%%%%%%%%%%%%%
\subsection{Dark count rate of the PMTs}
\label{sec:darkcounts}
%%%%%%%%%%%%%%%%%%%%%%%%%%%%%%%%%%%%%%%%%%%%%%%%%%

For checking stability of the detector such as dark count rate and DC baseline level of all the PMT channels, periodic trigger signals are generated with a waveform generator at low frequencies. Data are taken by default with the ``physics'' trigger, which is created from the signals from the PMTs, and the generator trigger, simultaneously. Events taken with the generator trigger are tagged in the data so that they can be identified in offline analysis. Currently, the frequency of the generator trigger is set to 20 Hz, which leads to approximately 1\% of the recorded events. 

\begin{figure}[hbtp]
\begin{center}
\includegraphics[width=0.6\columnwidth]{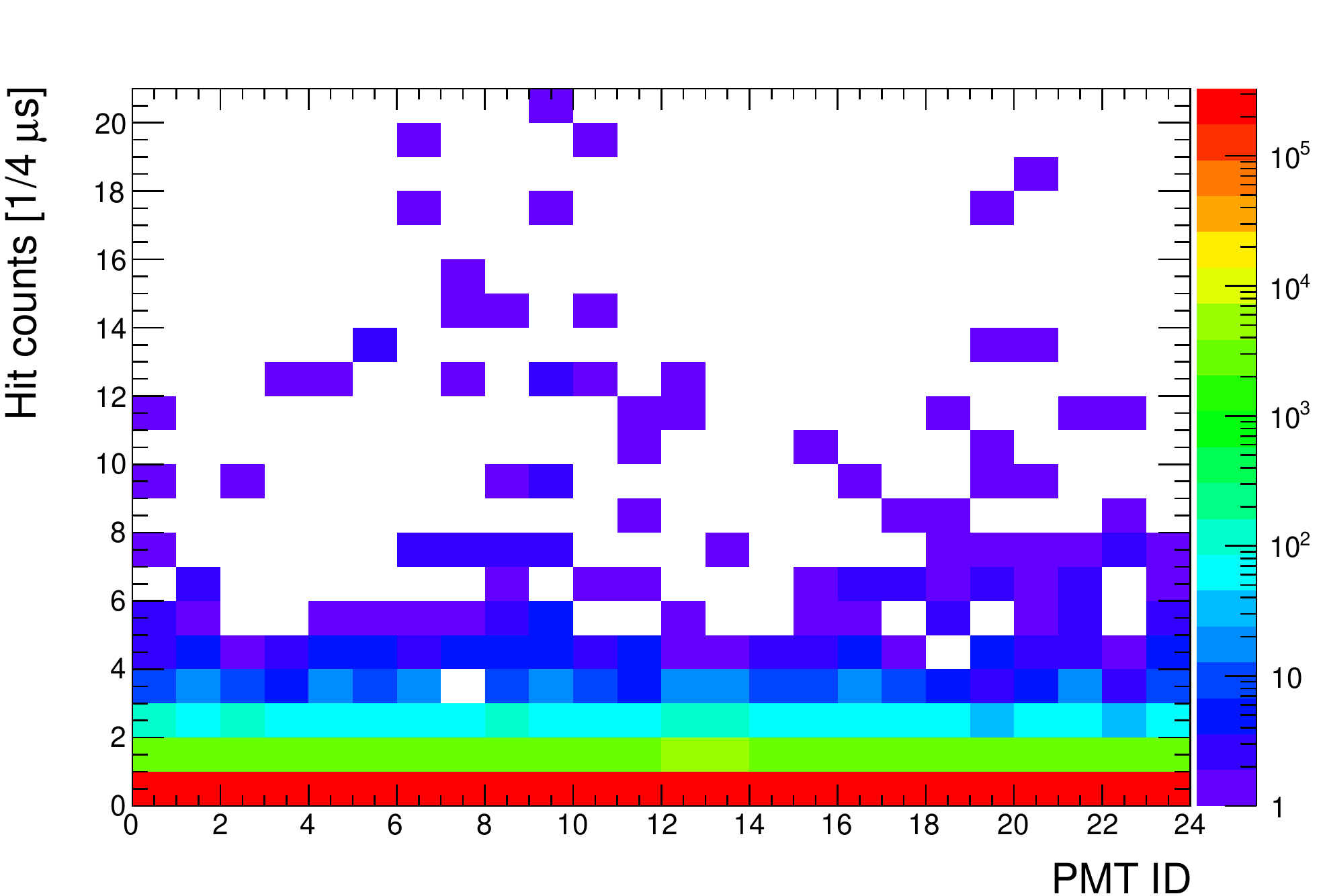}
\caption{Distribution of dark counts in the 4-$\mu$s acquisition window for each PMT.}
\label{fig:darkcounts}
\end{center}
\end{figure}

Dark count rate of each PMT was evaluated. 
After rejecting events contaminated from random coincidences of argon scintillation signals, hit counts distribution within the acquisition window of $\sim$\,4 $\mu$s was obtained as shown in Fig.\,\ref{fig:darkcounts}. 
Fitting Poisson distribution function to these distributions, 
the dark count rate for each individual PMT is 2.3 kHz on average per PMT. 
This rate corresponds to 0.0089 counts on average per PMT, i.e.~0.21 counts for the sum of 24 PMT channels in the acquisition window. 
Contribution from the dark counts to the integration of the total detected light (several tens of photoelectrons) can thus be considered as negligible.

%%%%%%%%%%%%%%%%%%%%%%%%%%%%%%%%%%%%%%%%%%%%%%%%%%
%\subsection{Data Analysis - Data Quality Cuts and $^{39}$Ar Spectrum}
%\subsection{Data quality study}
\subsection{Trigger threshold}
\label{sec:dq}
%%%%%%%%%%%%%%%%%%%%%%%%%%%%%%%%%%%%%%%%%%%%%%%%%%

The low dark count rates of the PMTs allow for the use of an OR trigger configuration. The signals from 12 PMTs in each array are summed using an analog fan-in/out and are discriminated to create logic signals with a threshold of $\approx$\,2\,pe for each array. The final trigger signal is generated either by a logic AND or OR of the two discriminator outputs. Switching from AND to OR trigger configuration increases the trigger rate by $\sim$\,30\%. 
\begin{figure}[htb]
\begin{center}
\includegraphics[width=0.6\columnwidth]{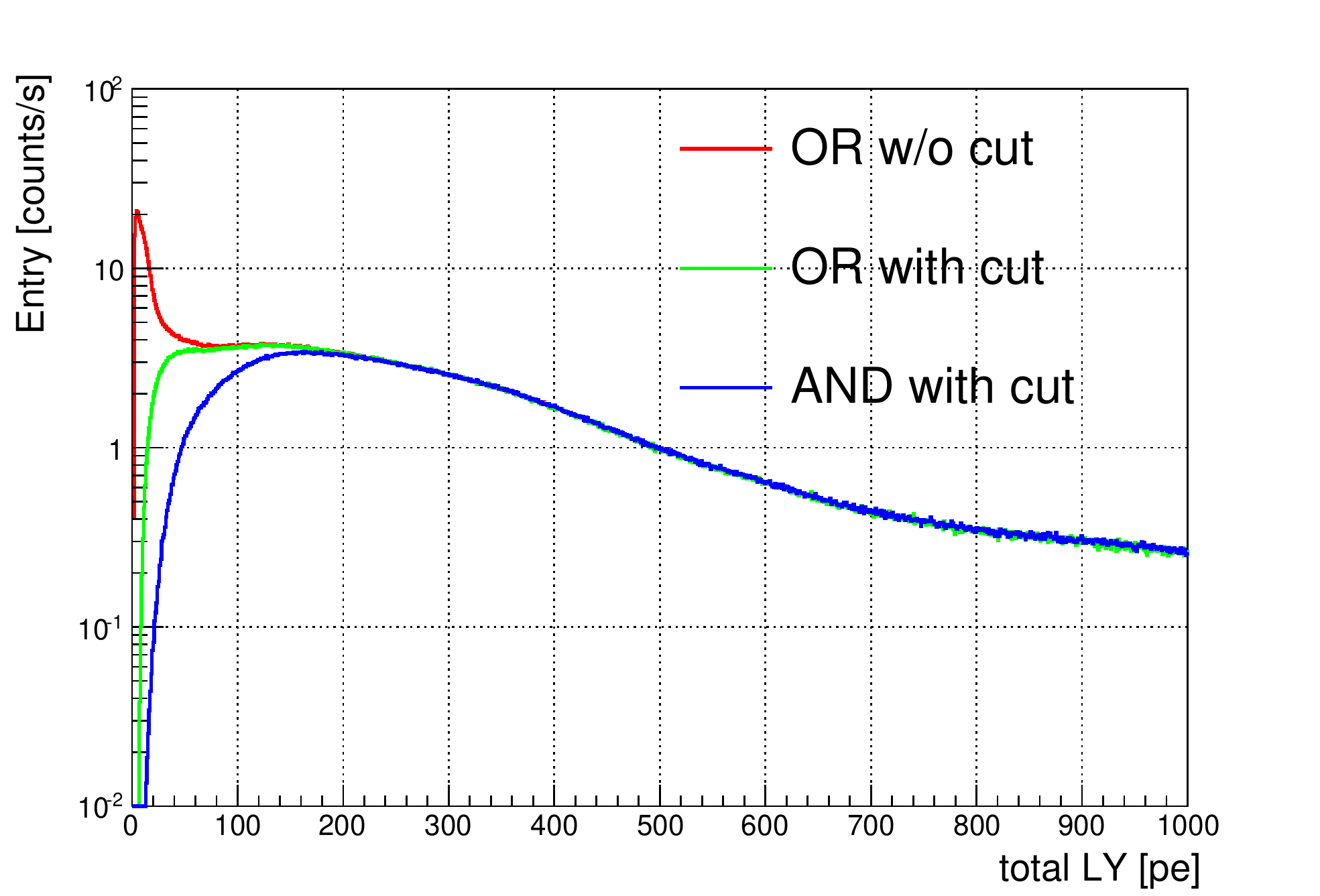}
\caption{Energy spectra obtained with OR (red) and AND (blue) trigger. Preliminary data quality cuts suppress efficiently the noise contribution in the OR trigger events, yet preserving relatively low effective threshold (green).}
\label{fig:triggers}
\end{center}
\end{figure}
Figure\,\ref{fig:triggers} shows the low energy edge of the energy spectra obtained with OR (red) and AND (blue) trigger. The sharp peak seen in the OR trigger events below $\sim$\,40 p.e.~may largely be attributed to noise. However, it can clearly be seen that OR trigger eliminates the threshold effect seen in the AND trigger events. 

\begin{figure}[htb]
\centering
\includegraphics[width=0.95\textwidth]{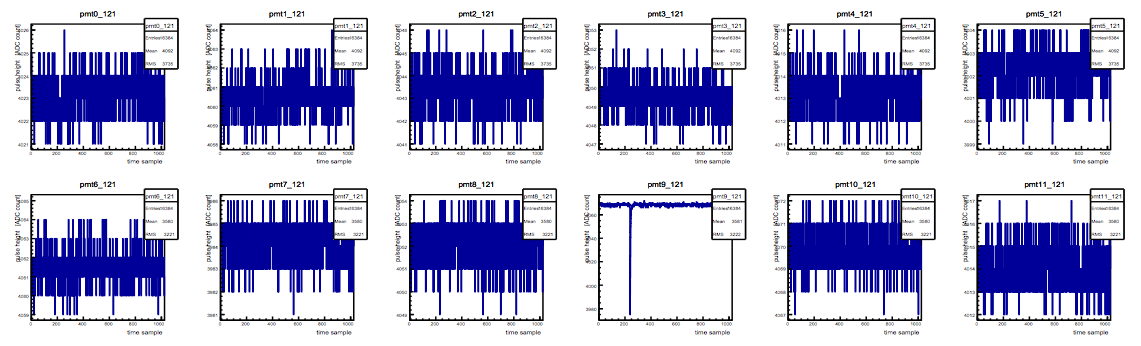}
\caption{Display of an event where the entire light is seen by one PMT only. Shown are the signals of the PMTs for the top array, the bottom array is completely empty. }
\label{fig:evtDisplay}
\end{figure}
In OR trigger configuration, we observed events where light is detected by one single PMT. Such an event is displayed in Fig.\,\ref{fig:evtDisplay} for PMT\,9 while the other PMTs of the top array and bottom array (not shown) exhibit only noise. These ``flashers'' can be discarded by simple cuts in the data analysis. Preliminary cuts were applied to the OR trigger events and, as can be seen in the green curve in Fig.\,\ref{fig:triggers}, the low energy peak is efficiently suppressed, yet preserving low effective thresholds. 

In March 2015 the thermal insulation at the top flange, made out of Perlite bricks, was replaced with sheets of extruded polystyrene after observing gamma-like background concentrated in the top section of the LAr target. Perlite is an amorphous volcanic glass and was found to be slightly radioactive mainly due to the presence of $^{40}$K. The removal of the perlite bricks reduced the gamma-like background at the top of the detector significantly resulting in a general decrease of the trigger rate by roughly 20\%. 

\subsection{Preliminary results from the full liquid argon target}

\begin{figure}[hbtp]
\begin{center}
\includegraphics[width=0.7\textwidth]{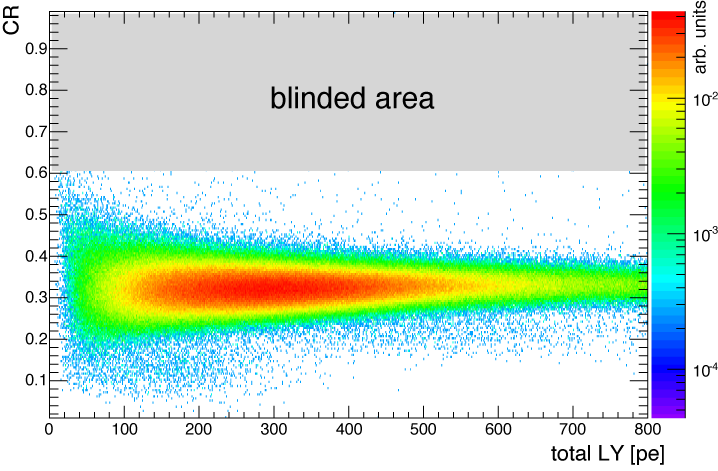}
\caption{\ardm\ data in a representation of the CR vs the total light yield (preliminary).}
\label{fig:e-recoil}
\end{center}
\end{figure}

The primary focus is the understanding of electron-like backgrounds and to investigate pulse shape discrimination (PSD). Different types of blind analyses, e.g. based on data stratification are
undertaken. Figure\,\ref{fig:e-recoil} shows preliminary data in a representation of component ratio (CR) versus total light (in pe). CR is calculated by the ratio of light recorded in the first 90~ns to the total for each event. 

Very few cuts were applied to select the data. The plot is dominated by the presence of $^{39}$Ar populating the band of electronic recoils situated around a value of 0.3 for CR. The obvious narrowness of the band is a first indication at the ton scale for the high power of pulse shape discrimination of LAr detectors. 
Nuclear recoils from neutrons or WIMPs are expected at values around around 0.8\,\cite{regenfus}. 

The signal acceptance band will be determined from data taken with external neutron sources. Recently a test run with neutrons was undertaken by applying a $^{252}$Cf source to the side of the \ardm\ main vessel. The $^{252}$Cf source used in this test emitted isotropically about 600 neutrons/s from spontaneous nuclear fission and  produces a continuous spectrum of fast neutrons up to several MeV.  Signals from neutrons could clearly be identified in the region of large values for CR. A more detailed analysis is ongoing and will be presented elsewhere.

%%%%%%%%%%%%%%%%%%%%%%%%%%%%%%%%%%%%%%%%%%%%%%%%%%
\subsection{Electronic background description by the Monte Carlo framework}
\label{sec:mc}
%%%%%%%%%%%%%%%%%%%%%%%%%%%%%%%%%%%%%%%%%%%%%%%%%%

To evaluate the quantity of electronic recoil background and to investigate its rejection factor by PSD we aim at a description of the $\beta$-spectrum generated by the presence of $^{39}$Ar in the target. The theoretical shape of the distribution is given by the following relation,

\begin{equation}
\frac{dN}{dw} \sim p w (w_0-w)^2 F(Z,w) S(w)
\label{equ:39ar}
\end{equation}
\vspace{2mm}

where $p$ is the $\beta$ momentum, $w = (m_{\rm e}c^2 + E^{\rm kin}_{\rm e})/m_{\rm e}c^2$, $w_0 = w + E_{\bar{\nu}}/m_{\rm e}c^2$, $Z$ the atomic number, $F(Z,w)$ the Fermi correction and $S(w)$ is a correction function for the forbiddenness of the transition (first forbidden Gamow-Teller transition), taken from references\,\cite{HDaniel,Segre,arXiv,beta}. The Q-value of the $^{39}$Ar decay is 565\,keV\,\cite{1950:brosi}.\\ 

\begin{figure}[h]
\centering
\includegraphics[width=0.49\textwidth]{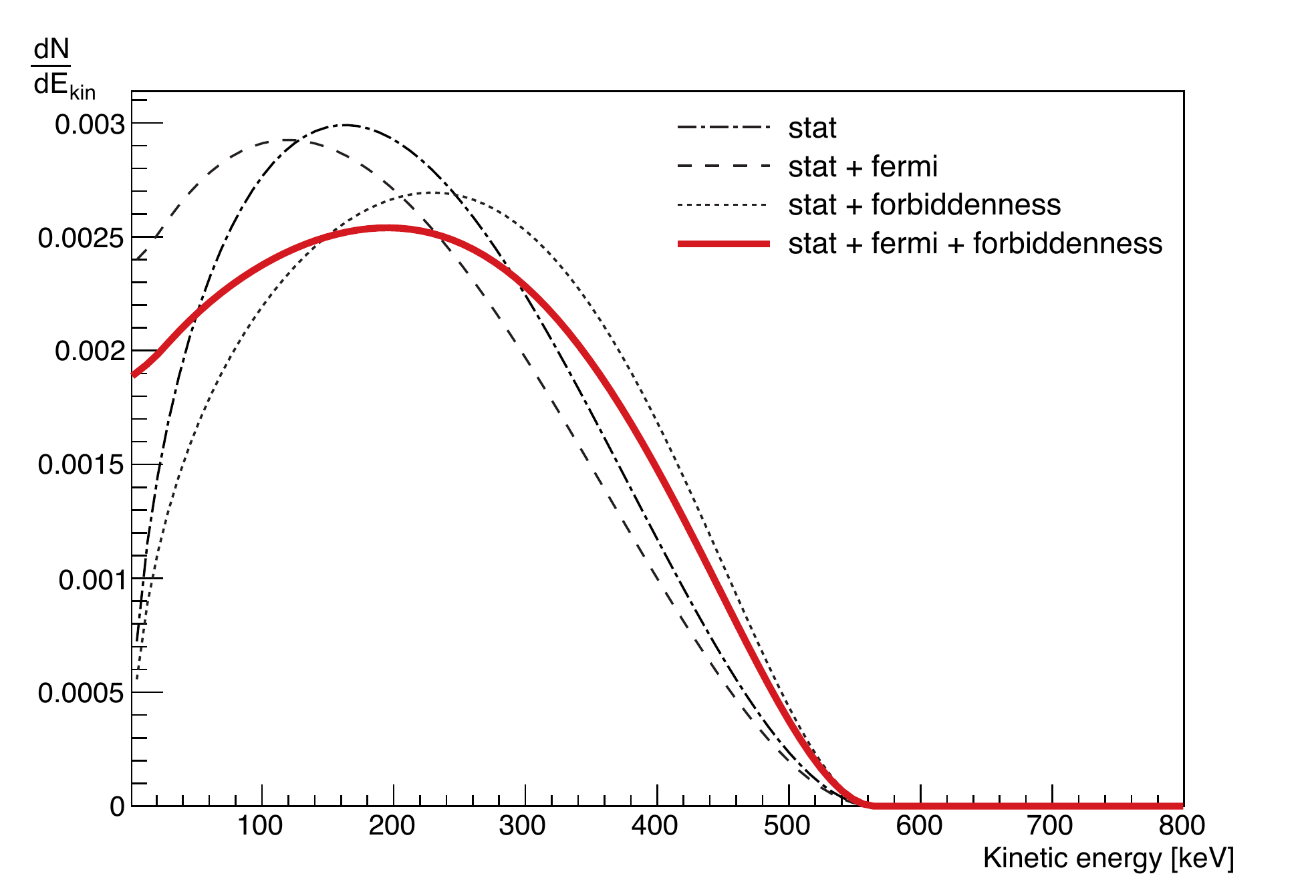}
\includegraphics[width=0.49\textwidth]{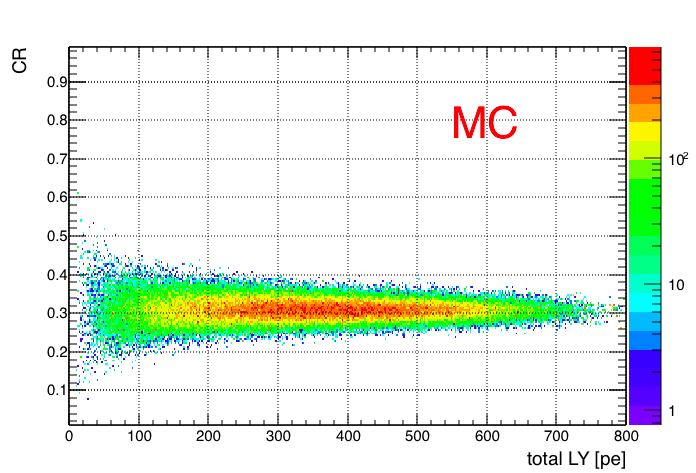}
\caption{Left: Theoretical spectrum (red curve) of the $^{39}$Ar $\beta$-decay with a Q-value of 565 keV. The dashed and dotted lines illustrate the effect of the corrections for the Fermi factor and the first unique forbidden transition (Gamow-Teller) to purely phase space distributed final states. Right: Component ratio (CR) vs.~total detected light in photoelectrons (pe) obtained from a preliminary Monte Carlo simulation of $^{39}$Ar events in the \ardm\ detector.} 
\label{fig:MC_Data}
\end{figure}

The red curve in Fig.\,\ref{fig:MC_Data} (left) displays the $\beta$-spectrum calculated from (\ref{equ:39ar}). The dashed lines show purely phase space distributed final states (stat), as well as their corrections by the Fermi term and the forbiddenness from the first Gamow-Teller transition. Based on this spectrum we generate $^{39}$Ar decays isotropically distributed over the entire liquid target 
with a Geant4 based simulation framework. The Monte Carlo (MC) code includes the description of all primary physics phenomena for particle interactions and signal production, as well as the subsequent optical processes as waveshifting and light propagation up to the generation of the PMT waveform output in the same format as used for real data. Figure\,\ref{fig:MC_Data} (right) 
shows preliminary MC data in the same representation as in Fig.\,\ref{fig:e-recoil}, of CR in function of the total number of (detected) photoelectrons. The MC data was treated by the same analysis software as was the experimental data. The results obtained by the comparison of real and simulated data are very promising and give us strong confidence in the understanding of the detector. 

%%%%%%%%%%%%%%%%%%%%%%%%%%%%%%%%%%%%%%%%%%%%%%%%%%%
%\section{Environmental neutron flux measurements in Hall A of LSC}
%\label{sec:in-situ}
%%%%%%%%%%%%%%%%%%%%%%%%%%%%%%%%%%%%%%%%%%%%%%%%%%%
%\input{Section_neutrons_v2}

%%\vspace{4mm}

%\section{Environmental Neutron Measurement Updates}
\section{Environmental neutron flux measurement in Hall A}

The \ardm\ Collaboration is carrying out a dedicated neutron measurement campaign 
in Hall A of the LSC Lab, in collaboration with the Nuclear Innovation Unit from CIEMAT,  in order to assess the \begin{figure}[h]
\begin{center}
\includegraphics[width=10.5cm]{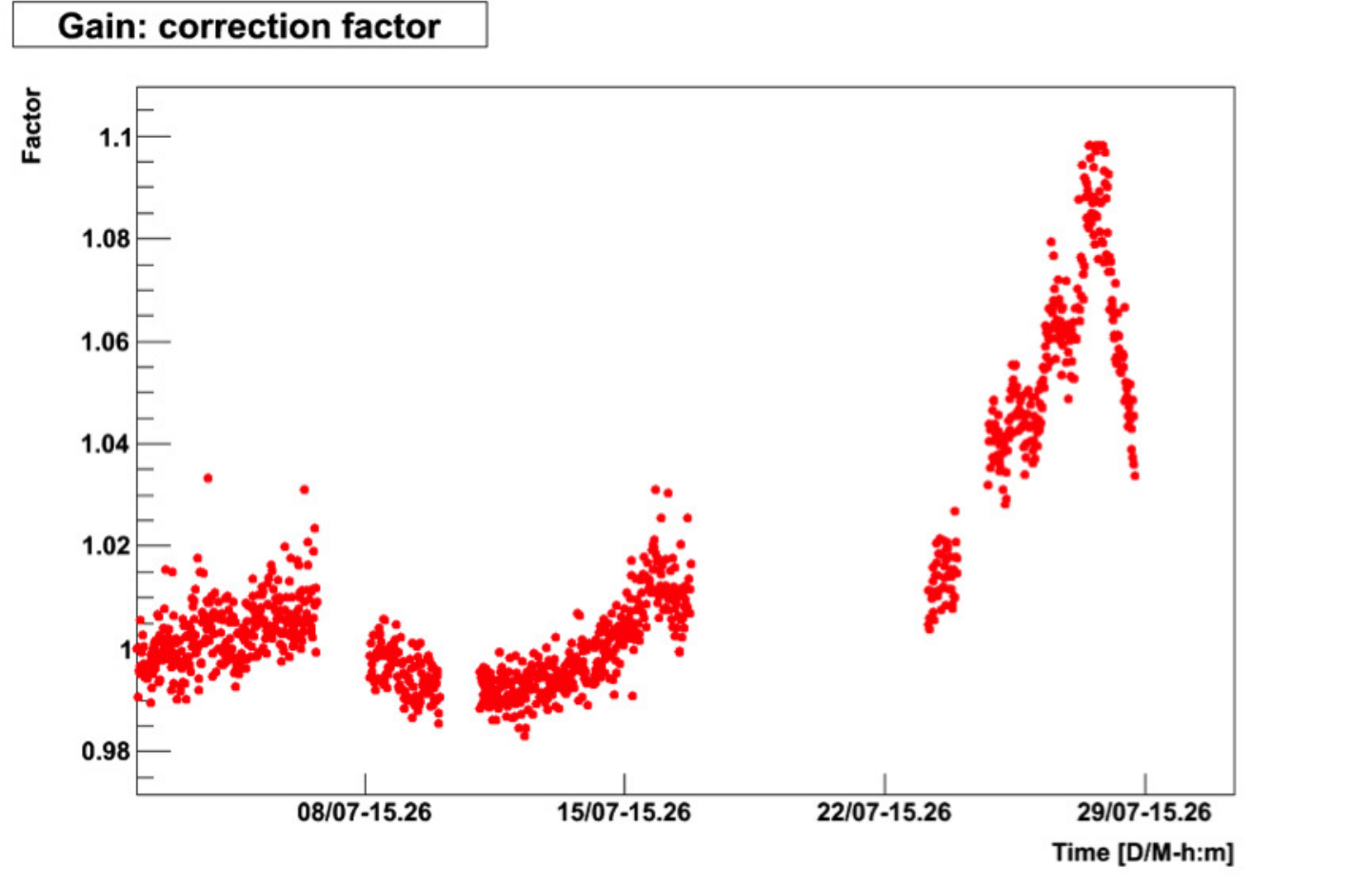}
\caption{Gain correction factor calculated through the drift  of the  $^{208}$Tl line during one month of data taking. }
\label{fig:Gain_correction}
\end{center}
\end{figure}
environmental neutron flux and estimate the possible neutron induced background relevant for the Dark Matter search. A BC501A liquid scintillator detector, for fast neutron spectroscopy, was installed underground in Hall A and took data continuously during the last year. After the initial commissioning of detector and DAQ underground (February--March 2014) we started the physics run in April 2014. 

The energy threshold was set to 100 keV$_{\rm ee}$, which allows to measure neutrons with energies $E \geq 1$ MeV, reducing at the same time the total counting rate. A total live time of 216 days have been obtained until March 2015, corresponding to more than 65\% of the total time, and  a total of $2.57\cdot10^9$ events have been \begin{figure}[h]
\begin{center}
\includegraphics[width=8cm]{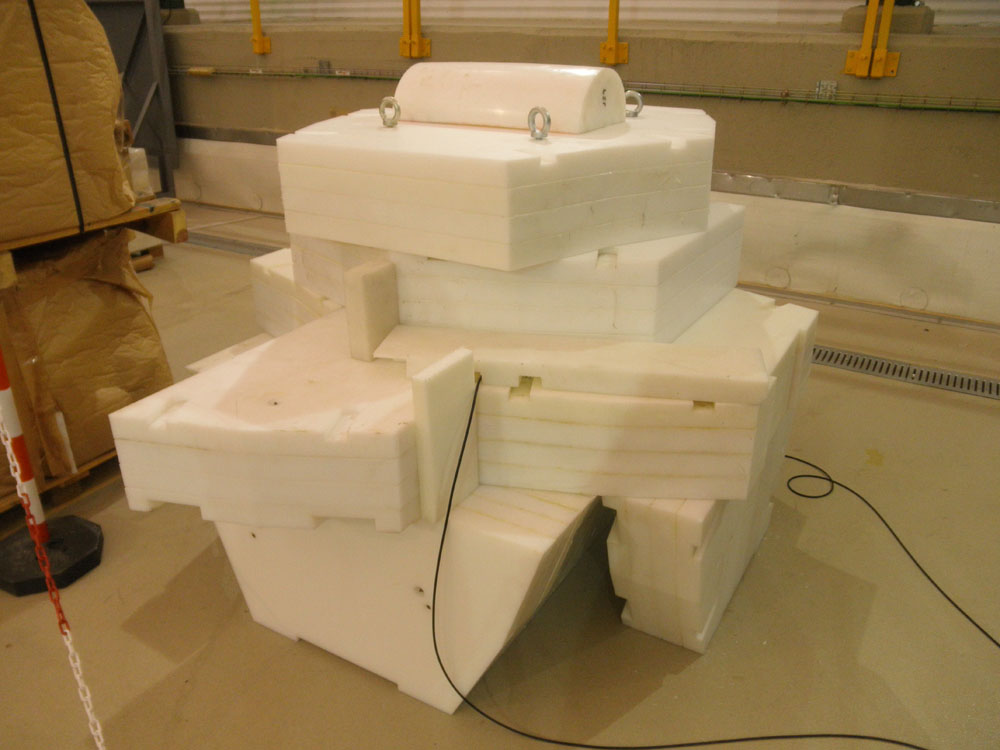}
\caption{Polyethylene castle constructed with spare polyethylene bricks from \ardm\ to measure the intrinsic background of the liquid scintillator detector.}
\label{fig:Poly_castle}
\end{center}
\end{figure}
recorded. Calibration runs have been taken periodically with the help of the LSC personnel with $\gamma$ sources ($^{22}$Na and $^{137}$Cs) in order to monitor the detector stability.
A preliminary tuning of the algorithms necessary for the neutron signal identification and the gamma background rejection was obtained. Some variations of the detector response ($\approx$10\% maximum), possibly driven by environmental parameters changes in the Hall A, have been observed during the data taking. Detailed investigations to correlate the changes in the detector response to parameters such as temperature and humidity are ongoing.  Different strategies to correct for the effect are also developed,  based on the variation in time of the position of the $^{40}$K and  $^{208}$Tl lines that are clearly visible in the gamma background spectra.  In Fig.\,\ref{fig:Gain_correction} we show the calculated gain correction factor for the data taken during July 2014. The study of the impact of such instabilities on the neutron flux measurements is in progress.

In March 2015 the detector was moved into a small neutron shielding (castle) built by spare polyethylene bricks from the \ardm\ external neutron shield (Fig.\,\ref{fig:Poly_castle}), with the goal of measuring the intrinsic background of the detector. The measurements should allow disentangling the contribution of the neutrons produced inside the detector by spontaneous fission and ($\alpha$,n) reactions from the measured neutron flux.

%%%%%%%%%%%%%%%%%%%%%%%%%%%%%%%%%%%%%%%%%%%%%%%%%%
\section{Field cage upgrade}
\label{sec:plan}
%%%%%%%%%%%%%%%%%%%%%%%%%%%%%%%%%%%%%%%%%%%%%%%%%%

Currently the experiment operates in physics data taking single-phase mode (\ardm\,Run\,I). Our goal is to keep the detector under stable operating conditions for about three months to accumulate a large amount of data to study the background with $\geq$\,10$^{9}$ events, and to explore the signal region with neutron sources. Periodic runs with internal and external calibration sources will be done to monitor the operational conditions of the detector.
A ``long shutdown" period is planned for summer 2015 to insert a new TPC field cage, replace the side reflectors and to insert some prototype light readout modules based on SiPMs. 

\begin{figure}[h]
\begin{center}
\includegraphics[width=0.6\textwidth]{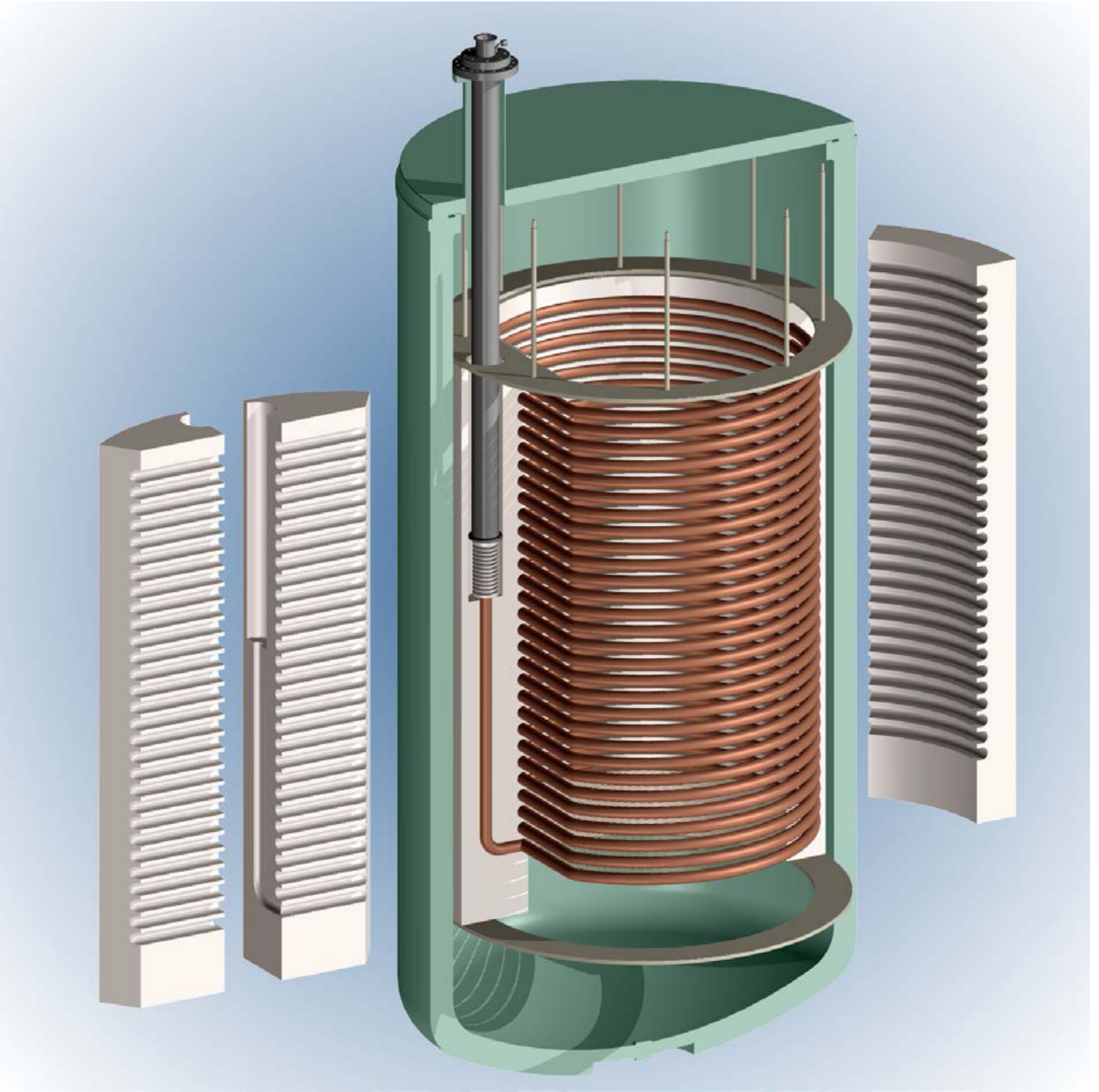}
\caption{New field cage with %boronated 
borated PE shielding, ready to be installed into ArDM.}
\label{fig:FieldCage}
\end{center}
\end{figure}

The main reason for the experimental shutdown is the installation of the TPC field cage. 
The new field cage contains additional (borated) polyethylene (PE) shielding to suppress neutrons from detector components and is shown in Fig.\,\ref{fig:FieldCage}. The PE pieces also serve as protection against discharges and 
%\begin{wrapfigure}{r}{0.6\textwidth}
%\vspace{-4mm}
%\centerline{\includegraphics[width=0.6\textwidth]{picts/NewfieldCage.pdf}}
%\vspace{-4mm}
%\caption{New field cage with boronated PE shielding, ready to be installed into ArDM.}
%\label{fig:FieldCage}
%\vspace{-6mm}
%\end{wrapfigure}
keep the HV lines embedded as much as possible. All components for the new field cage have been produced, are delivered to LSC and are ready to be installed. The corresponding high voltage system including the feedthrough was successfully tested at CERN to withstand voltages up to 150 kV, while less than 100 kV is needed to run \ardm. At the time of  the installation of the new field cage we also plan to replace the present side reflectors with a new set of non metallic, highly  specular reflecting foils (3M Vikuiti ESR foils) coated with a thin layer of TPB. For this purpose we are setting up a large area evaporator at our laboratories at CIEMAT. 

\section{R\&D towards a light readout with SiPMs}
\label{sec:up}

The light readout system is among the most important instrumentation for the \ardm\ detector and determines crucially its performance in terms of detection threshold and power for pulse shape discrimination.
The present light detection system based on PMTs is working according to its specification. However overall performance is limited due to the large amount of glass and the reduced quantum efficiency of the large area PMTs ($\sim$\,16\%) due to the platinum underlay necessary for their cryogenic operation. Several scenarios for upgrading the light readout in \ardm\ were considered over the last years. With recent progress in the development of silicon photomultipliers (SiPM) a promising alternative to PMTs now seems to be on the market. 

Since the technology is still very new we started an extensive R\&D program in the ETHZ group at CERN. The aim is to build a couple of prototype modules equipped with large area SiPMs capable to work at the temperature of LAr and to detect the time 
resolved scintillation light with the help of the wavelength shifter deposited on to a glass plate in front of the sensors. Hardware efforts are supported by a full optical simulation with the parameters of \ardm. 

\begin{figure}[htb]
\begin{center}
\includegraphics[width=0.6\textwidth]{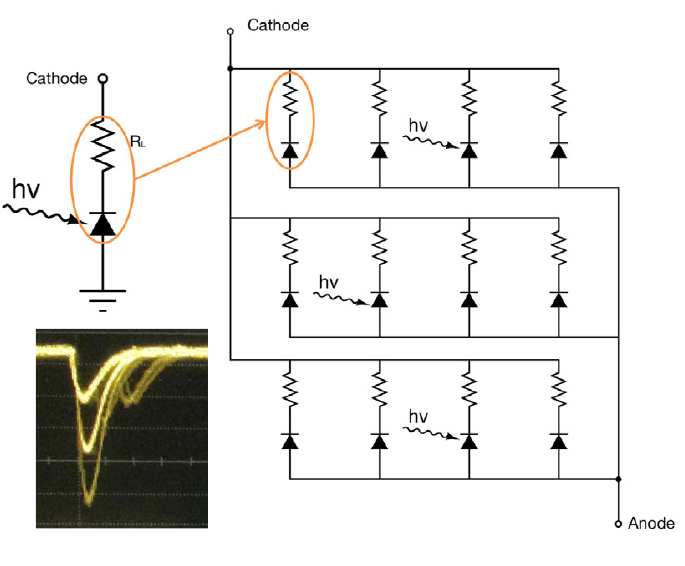}
\caption{Schematic diagram of a SiPM. One microcell contains a quenching resistor and a small avalanche photodiode operating in Geiger mode. The inset shows signals from a 7$\times$7\,mm$^{2}$ test sensor at LN$_{2}$ temperature.}
\label{fig:SiPMscheme}
\end{center}
\end{figure}

SiPMs work on the principle of an array of small cells (typ.~20\,$\mu$m) of avalanche photodiodes operating in Geiger mode (see Fig.\,\ref{fig:SiPMscheme}). Due to the very precise modern manufacturing methods the spread in the performance of individual cells allows for a common bias (typ. 25--70 V) with a large margin to drive the devices into a smooth breakdown (tolerances for overbiasing are typ.~5\,V). This property is crucial for their low temperature behaviour since the working points of all cells shift homogeneously to guarantee stable working operation.

Since a few years SiPM are produced at much larger sizes (up to 1\,cm$^{2}$) suited to build square-meter photo detector planes. They combine advantages of silicon photodiodes (fast, high QE) with an almost binary readout since the signal is produced by the defined charge stored per microcell (typ.~10$^{6} q_{\rm e}$). If several photons are detected simultaneously the output charge is a multiple of one 
%\begin{wrapfigure}{r}{0.7\textwidth}
%\vspace{-1mm}
%\centerline{\includegraphics[width=0.7\textwidth]{picts/SiPMphotoTSV.pdf}}
%\vspace{-1mm}
%\caption{Example of the latest generation TSV-packaged SiPM sensor from the company SensL, showing the almost borderless top surface (left) and backside with ball bumps (right). The sensor is composed by the actual SiPM (70\,$\mu$m) and a glass substrate of 320\,$\mu$m thickness.}
%\label{fig:SiPMphotoTSV}
%\vspace{-1mm}
%\end{wrapfigure}
cell charge leading to unprecedented photon counting capabilities (Fig.\,\ref{fig:SiPMscheme} inset). The limiting factor of the spread is determined by the noise of the connected preamplifier necessary to read the signal against the electric capacitance which can be substantial for large area SiPMs (several nF) since it is created by the parallel connection of all cells. This also leads to a limitation in rise time of the signal. Further on quenching resistors, responsible for the recharge of the cells, increase their value at lower temperatures in some of the devices leading to a long signal tail (up to $\mu$s) which can destroy the fast response necessary for the reconstruction of the pulse shape in LAr.

However due to a cut by half of the thermal generation of electron hole pairs in silicon for every 10\,K decrease in temperature, the dark current (= dark count rate) almost disappears at the temperature of LAr. Assuming a dark count rate of 10\,Hz/cm$^{2}$ (measured on a single device in the lab) at 77\,K, a light readout system covering the entire readout areas of a detector like \ardm\ ($\sim$\,1\,m$^{2}$) will have a probability of 0.1 noise counts (pe) in a 1-$\mu$s integration window. Our present PMT system ($24\times 8$'' PMTs) has about half of this value ($\sim$\,50 kHz total dark count rate). This fact is a crucial  design parameter for large photon counting SiPM arrays.    

%\vspace{5mm}\noindent{\bf SensL sensors and construction of arrays}\\
\subsubsection*{SensL sensors and construction of arrays}
\begin{figure}[h]
\begin{center}
\includegraphics[width=0.7\textwidth]{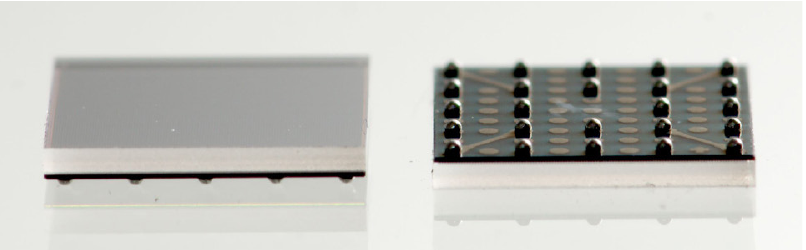}
\caption{Example of the latest generation TSV-packaged SiPM sensor from the company SensL, showing the almost borderless top surface (left) and backside with ball bumps (right). The sensor is composed by the actual SiPM (70 $\mu$m) and a glass substrate of 320 $\mu$m thickness.}
\label{fig:SiPMphotoTSV}
\end{center}
\end{figure}

%\noindent 
Considering different manufacturers in the first place now we decided to use the latest TSV-type model (Fig.\,\ref{fig:SiPMphotoTSV}) from the company SensL\,\cite{sensl} for the first prototypes. The decision was based on the optical 
%\begin{wrapfigure}{l}{0.35\textwidth}
%\vspace{-5mm}
%\centerline{\includegraphics[width=0.35\textwidth]{picts/SiPMarray2.pdf}}
%\vspace{-4mm}
%\caption{Array of SiPMs, older types have large 
%dead space ($\sim$\,2 mm) in between the sensors.}
%\label{fig:SiPMarray}
%\vspace{-2mm}
%\end{wrapfigure}
and mechanical performance of the device, their radio purity\,\cite{nextscreening}, as well as on their availability (and price) for large scale production\footnote{SensL is using a standard CMOS process which can be transferred to several different semiconductor fabrication plants (FABs) if necessary.}.  
These 3rd generation sensors (internally named as J-Series\,\cite{sensl}) feature photon detection efficiencies (PDE) 
of $>$50\% in the emission range of our wavelength shifter, as well as tiling possibilities with as little dead space as 100 $\mu$m in-between. The value for PDE contains the intrinsically high value for QE in Si ($\sim$\,70\%) multiplied with fill factor F ($\sim$\,80\%) given by dead areas between the cells and on the border of the sensor. Figure\,\ref{fig:SiPMarray} shows the example of an array still of older type sensors leading to an overall fill factor of below 60\%, while values around 80\% can be achieved with the TSV type sensors. This new generation also features dark-count and after-pulsing rates reduced by a factor of 20 towards the older models. Improvements have also been made to both the standard (anode-cathode) rise time and the cell recovery time. J-Series sensors are available in a $6\times 6$ mm$^{2}$ active area size and are compatible with an industry standard, lead-free, reflow soldering processes. This, together with the small amount, and the cleanliness of the used materials (silicon and glass\,\cite{SiO2forGlass}) opens the way for the production of ultra low background photon detectors.

%
%\vspace{5mm}\noindent{\bf Towards a design for a SiPM based readout in ArDM}\\
\subsubsection*{Towards a design for a SiPM based readout in ArDM}
%\noindent 
%\begin{wrapfigure}{l}{0.42\textwidth}
%\vspace{-1mm}
%\centerline{\includegraphics[width=0.42\textwidth]{picts/SiPMTsArDM37.pdf}}
%\vspace{-1mm}
%\caption{First proposal of a possible arrangement of 37 tiles equipped with each 49 SiPMs read out by a x-y matrix.}
%\label{fig:SiPMTsArDM37}
%\vspace{-1mm}
%\end{wrapfigure}
The upgrade of the light readout is planned to be done in two main steps. In a first step we plan to replace some of the lower central PMTs with a number of SiPM arrays during the shutdown scheduled for the insertion of the TPC field cage. Existing readout lines and FADC channels will be used. The most important parameters such as light \begin{figure}[h]
\centering
\begin{minipage}{.5\textwidth}
  \centering
\includegraphics[width=0.92\textwidth]{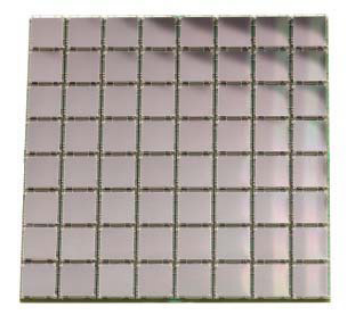}
\caption{Array of SiPMs, older types have large dead space ($\sim$\,2\,mm) in 
between the sensors.}
\label{fig:SiPMarray}
\end{minipage}%
\begin{minipage}{.5\textwidth}
  \centering
\includegraphics[width=0.8\textwidth]{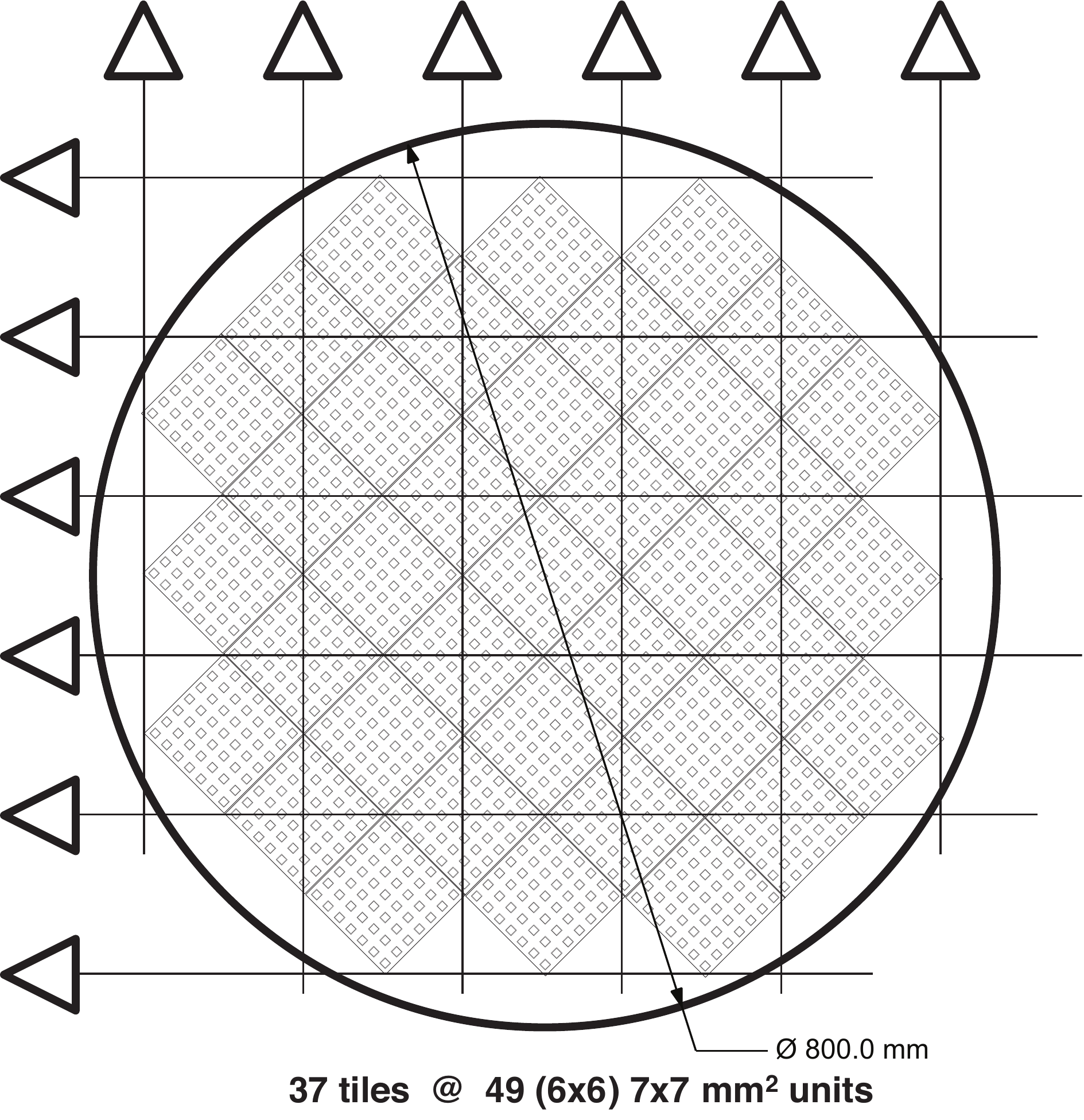}
\caption{First proposal of a possible arrangement of 37 tiles equipped with each 49 SiPMs read out by a x-y matrix.}
\label{fig:SiPMTsArDM37}
\end{minipage}
\end{figure}
yield, dark count rates and signal bandwidth will be verified. In parallel to these measurements and experimental tests we are developing a detailed simulation package for the entire SiPM readout system, separately done in Spice for the electronic circuits, as well as in Geant4 for the light collection. With results from the test modules we plan to converge on the final design for a readout plane. A possible first layout of 37 tiles with each 49 SiPMs is shown in Fig.\,\ref{fig:SiPMTsArDM37} on the example of 12 readout lines in a 2D arrangement. For better performance one could increase the number of readout channels (e.g. 32 per plane). The optimal level of multiplexing will be defined when the final module design is known to cope with true signal shapes, dark count rates as well as the photon occupancy and the necessary position reconstruction of the detector (simulations). 

%%%%%%%%%%%%%%%%%%%%%%%%%%%%%%%%%%%%%%%%%%%%%%%%%%
\section{Conclusions and outlook}
\label{sec:conclusions}
%%%%%%%%%%%%%%%%%%%%%%%%%%%%%%%%%%%%%%%%%%%%%%%%%%

In February 2015 the \ardm\ experiment achieved a major milestone by filling the detector with a total of roughly 2\,t of LAr, marking its transition to a period of physics data taking in the single-phase operation mode. Preliminary results confirm the dominance of decays of $^{39}$Ar isotopes present in the atmospheric argon. We presented first indications that background discrimination by pulse-shape-discrimination can be  efficiently achieved with ton-scale LAr detectors. The shape of the electronic-recoil band from real data is compatible with the distribution from data generated by the \ardm\ MC framework. With more than 10$^8$ events already taken at the full target a range of studies are now ongoing. 
The planning foresees to extend the first single-phase run (\ardm\, Run\,I) and to ensure precise calibration with external and internal radioactive sources. Our analysis program will concentrate on the understanding of the backgrounds and to investigate PSD. 
We expect to increase our data set by more than a factor of 10 by summer 2015, allowing for an unprecedented high statistics study of parameters relevant for LAr Dark Matter detectors. 

One year of continuous data taking has  been successfully accomplished in the neutron measurement campaign to assess the environmental neutron flux in Hall A of LSC with dedicated liquid scintillator and $^{3}$He detectors, resulting in a total live time of 216 days and more than 10$^9$ recorded events. Detailed analyses are currently being pursued, with special attention to systematic errors. %Presently the intrinsic background of the neutron detectors is determined by the use of a polyethylene shield. 

Starting from July 2015, a ''long shutdown'' is planned to install the TPC drift cage. The intervention will also be used to replace the side reflectors and to install some prototype light readout modules based on SiPMs. An extensive R\&D project has begun for the development of the SiPM-based light readout system. Such an R\&D project will extend towards the future evolution of \ardm\ and potentially also large-scale future LAr Dark matter detectors. 

%%%%%%%%%%%%%%%%%%%%%%%%%%%%%%%%%%%%%%%%%%%%%%%%%%%%
\section*{Acknowledgement} 
%%%%%%%%%%%%%%%%%%%%%%%%%%%%%%%%%%%%%%%%%%%%%%%%%%%%

The \ardm\ Collaboration thanks the Laboratorio Subterr\'{a}neo de Canfranc and its staff, as well as the Scientific Committee of LSC for their continuous support. We acknowledge the kind support of the NEXT Collaboration by making available the evaporator for coating of the side reflectors. We are grateful to The Nuclear Innovation Unit at CIEMAT and the CUNA Collaboration for their participation and support in the neutron background measurements campaign.

%%%%%%%%%%%%%%%%%%%%%%%%%%%%%%%%%%%%%%%%%%%%%%%%%%


\begin{thebibliography}{99}
\bibitem{SiO2forGlass}
The (standard) glass presently used can be replaced by synthetic fused silica which has excellent 
radio purity properties (private communication with the producer).
\end{thebibliography}
\end{document}